# Externally Occulted Terrestrial Planet Finder Coronagraph: Simulations and Sensitivities

Richard G. Lyon[a], Sally Heap[a], Amy Lo[b], Webster Cash[c], Glenn D. Starkman[d], Robert J. Vanderbei[e], N. Jeremy Kasdin[e], Craig J. Copi[d]

[a]NASA/Goddard Space Flight Center, Greenbelt, MD;
[b]Northrop Grumman, Los Angeles, CA; [c]University of Colorado, Boulder, CO; [d]Case Western Reserve University, Cleveland, OH; [e]Princeton University, Princeton, NJ

## ABSTRACT

A multitude of coronagraphic techniques for the space-based direct detection and characterization of exo-solar terrestrial planets are actively being pursued by the astronomical community. Typical coronagraphs have internal shaped focal plane and/or pupil plane occulting masks which block and/or diffract starlight thereby increasing the planet's contrast with respect to its parent star. Past studies have shown that any internal technique is limited by the ability to sense and control amplitude, phase (wavefront) and polarization to exquisite levels - necessitating stressing optical requirements. An alternative and promising technique is to place a starshade, i.e. external occulter, at some distance in front of the telescope. This starshade suppresses most of the starlight before entering the telescope - relaxing optical requirements to that of a more conventional telescope. While an old technique it has been recently been advanced by the recognition that circularly symmetric graded apodizers can be well approximated by shaped binary occulting masks. Indeed optimal shapes have been designed that can achieve smaller inner working angles than conventional coronagraphs and yet have high effective throughput allowing smaller aperture telescopes to achieve the same coronagraphic resolution and similar sensitivity as larger ones.

Herein we report on our ongoing modeling, simulation and optimization of external occulters and show sensitivity results with respect to number and shape errors of petals, spectral passband, accuracy of Fresnel propagation, and show results for both filled and segmented aperture telescopes and discuss acquisition and sensing of the occulter's location relative to the telescope.

## 1. INTRODUCTION

Direct detection and characterization of terrestrial (Earth-like) planets in orbit around nearby stars remains a tantalizing proposition. Planets are expected from a few tens to a few hundred milli-arcseconds in angular separation from nearby stars and of order $10^{-10}$ times dimmer in visible light, and likely embedded in an unknown sea of scattered light from dusk surrounding the star and seen through local dust in our solar system. A multitude of coronagraphic approaches for increasing planetary contrast relative to starlight, allowing for angular separation of the planet light from its star and ultimately spectroscopy of the planet, have been studied. These approaches generally consist of a single telescope with an internal starlight suppression scheme. Various schemes abound and consist of either shaped of apodized pupil masks, and/or focal plane masks, or complex shaped optics, which emulate apodization or an internal nulling interferometer. Each of these internal methods have differing yet extreme requirements on wavefront, amplitude and polarization generally requiring some form amplitude, wavefront, and polarization sensing and/or control with stressing optical and stability requirements. These stressing requirements are due to incompletely suppressed diffracted/scattered starlight leaking through to the focal plane subsequently reducing the contrast of the planet with respect to the suppressed starlight.

An alternative approach is to suppress the starlight prior to entering the telescope, thereby mitigating stressing requirements in the telescope system and relaxing it to that of more conventional space telescope technology. This is an old approach dating back to L. Spitzer [1] and can be realized in space by flying two spacecraft: one consisting of the telescope system and the other an external occulter. The external occulter, at distance z in front of the telescope (Fig-1) along the line of sight to the stellar system under study, creates a deep shadow and the telescope resides within it. The starlight is suppressed and the planet light, off-axis relative to the line of sight, passes the edge of the occulter and enters the telescope aperture without reduction in throughput and independent of wavelength. The *geometric* inner working angle (IWA) is the ½ angle subtended by the occulter as seen from the telescope, i.e. $\phi_{IWA} = W/2z$ where W is the diameter of the occulter and z the separation of the occulter-to-telescope, *geometric* since this is only true to the limit of geometric optics, i.e. as the wavelength tends to zero. In practice the *diffractive* IWA, defined herein as that angle at which the contrast of the planet to suppressed starlight exceeds unity, is slightly smaller than the geometric IWA due to diffraction and only a weak function of wavelength over the range of interest. The depth of suppression and focal plane contrast vary with wavelength, occulter width, separation and telescope aperture diameter.

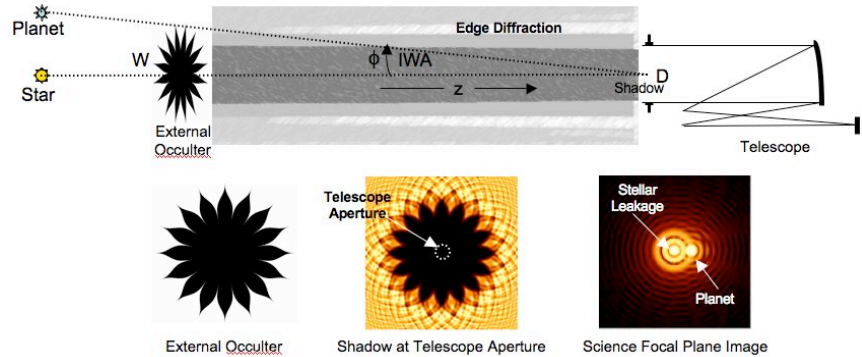

**Figure 1** – *Externally Occulted Coronagraph*

An external occulter typically contains hard edges, e.g. a circular disk, which causes Fresnel diffraction effects that tend to fill in or brighten the shadow thus leaking starlight through the telescope. Marchal [2] studied the use of large (200-800meter) petaled external occulters placed at separations of $10^5$-$10^6$ km for potential use with the Hubble Space Telescope (HST) for planetary detection but deduced it was infeasible for HST due to its orbital configuration. Copi and Starkman [3] studied more reasonably sized external occulters (~70 m) at separations of 50,000-100,000 km but with apodized transmission, i.e. graded transmission which is blocking in the center and changes continuously to transmitting towards it edge to better mitigate diffraction effects. Shultz et.al [4] studied a hybrid approach consisting of hard-edged (circular and square) occulters but coupled to an internal apodizer to theoretically obtain suppression levels of $10^{10}$ and Jordan et.al [5] conducted a ground demonstration using a square occulter to suppress Polaris. Cash [6] studied the use of a hyper-Gaussian apodization scheme but realized that a binary petaled occulter would well approximate an apodized occulter and subsequently demonstrated this approach in the lab to $\sim10^{-7}$ with broadband light. Vanderbei et.al [7] using constrained linear optimization designed an optimal 1D radial apodization function that suppresses broadband (0.4 – 1.2 microns) while simultaneously maintaining the intensity in the telescope aperture at $10^{-10}$ and approximated this apodized occulter using a binary petaled occulter. This Vanderbei form for the external occulter is currently the most effective design for an occulter which performs all the suppression external to the telescope, and has the flattest spectral response, and is studied in more detail herein under the NASA/Terrestrial Planet Finder Occulter (TPF-O) mission concept [8].

This approach does require two spacecraft [9] each with a spacecraft bus [10], attitude control, fuel and communications, and when acquiring a new target the telescope just re-points but the occulter must "fly" to the new line of sight to the target star, or alternatively both the telescope and occulter must reposition themselves. This levies additional requirements on the system and reduces the science duty cycle – however it may be the only viable approach with existing technology for direct planetary detection. This approach contains large design margins since suppression of starlight to $10^{-10}$ at the telescope aperture yields a focal plane contrast significantly higher at the IWA.

Hybrid approaches are also possible whereby an external occulter performs partial suppression but subsequently cascaded with an internal coronagraph within the telescope. This approach would require more stringent telescope tolerances - increasing the telescope's cost, but may allow a smaller closer in occulter with relaxed tolerances and lower fuel mass, while increasing the science time since less time is required to "fly" to the next target star. Thus a complex trade space exists between science, technical feasibility and cost. Studies of this complex trade space are actively being pursued. Herein we develop the occulter model, and model with errors and show parametric simulations with the end goal of ultimately exploring this complex trade space.

## 2. TERRESTRIAL PLANET FINDER OCCULTER

### 2.1 Baseline TPF-O Architectures

The TPF-O science requirements, delineated in [8], have resulted in two baseline TPF-O mission architectures (Table-1) for further study. These architectures consist of a single external 16 petal occulter coupled to either a 4 or 2.4 m telescope with 50 or 36.67 m occulter diameters at 72,0000 or 38,720 km with geometric IWAs of 72 and 98 mas respectively. The occulter's petaled shape is the Vanderbei form optimized over the spectral range of 0.4 – 1.0 microns. V-band (500 – 600 nm) aperture fluxes are as shown in Table-1 and for a planet at $10^{-10}$ would yield 0.13 and 0.045 planetary photons/sec for the 4.0 and 2.4 m apertures respectively. Assuming a product of the telescope optical transmission with the detector quantum efficiency of 0.5 yields reasonable times to photon limited signal-to-noise ratio (SNR) of 5 in V-band of 398 sec and 1105 seconds respectively. The time to SNR is purely the SNR without any exo- or local-Zodi nor assuming any leakage of the starlight and it is shown purely as a lower bound – actual integration times will be longer.

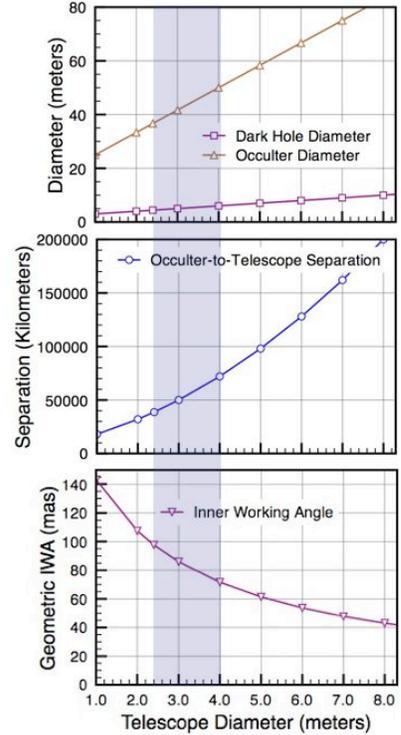

**Figure 2** – TPF-O *Scaling with Telescope Diameter*

**Table 1:** *TPF-O Baseline Architectures*

| Tel Diam (meters) | Separation Occ-to-Tel (kilometers) | Occ Diam Tip-to-Tip (meters) | Dark Hole Diameter (meters) | Geom IWA (mas) | $\lambda/D$ $\lambda = 0.5$ um (mas) | Geometric IWA ($\lambda/D$) | $M_v = 5$ Ap Flux (ph/sec) | $\Delta t$ to ph lim SNR=5 (seconds) |
|---|---|---|---|---|---|---|---|---|
| 4.0 | 72,000 | 50.00 | 6.0 | 72 | 25.78 | 2.78 | 1.3E+09 | 398 |
| 2.4 | 38,720 | 36.67 | 4.4 | 98 | 42.97 | 2.27 | 4.5E+08 | 1105 |

Scaling the as-designed occulter to other telescope sizes is straightforward since Fresnel diffraction scales with Fresnel number. The as-designed occulter operates at a Fresnel number of $W^2/\lambda z = 69.4$ at $\lambda$=0.5 microns for a 50 meter occulter at 72,000 km. If the Fresnel number is conserved, the functional form of the occulter's shadow remains unchanged, only its spatial scale is changed. This can be shown by a simple change of variables on the Fresnel integral (Eqn-1). Changing to an occulter of size 36.67 meters implies that $z_2 = (W_2/W_1)^2 z_1$ where $z_1$=72,000 km separation and $W_1$=50 m yields a separation of 38,720 km. The central dark hole diameter of 6.0 meters scales linearly with occulter size to 4.4 meters and the geometric IWA scales from 72 to 98 mas. Fig-2 plots these scaling relations where the shaded region is the range of a 2.4 – 4.0 meter aperture. The geometric IWA is the ½ angle subtended by the occulter diameter as seen from the telescope, i.e. $\phi_{IWA} = W/2z$ in the small angle approximation.

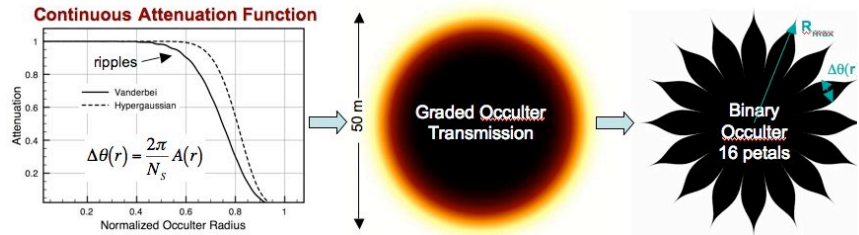

**Figure 3** – *Continuous Occulter to Binary Occulter*

### 2.2 Binary External Occulter

A radially symmetric occulter can be well approximated by a binary one. The left side of Fig-3 shows a hyper-Gaussian [6] and a Vanderbei [7] form of amplitude attenuation function $A(r)$ versus normalized radius. Fig-3 middle is the transmission of the occulter obtained by rotating the attenuation function into 2D about $r = 0$ and it is totally blocking in the center and continuously changing to complete transmission at the edges and beyond. Fig-3 right is the 16-petal binary approximation to the graded occulter. The binary occulter is obtained by requiring the integral of the transmission function around a radial zone to be the same for both the graded and binary occulter,

i.e. $\int_0^{2\pi} A_{graded}(r)d\theta = \int_0^{2\pi} A_{binary}(r,\theta)d\theta$ for any r

where $A_{graded}(r)$ is plotted on the left side of

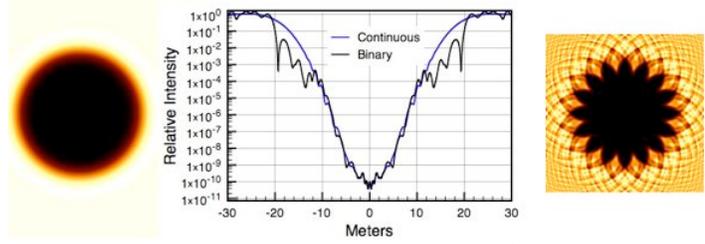

**Figure 4** - *Intensity at Aperture Plane of Telescope*

Fig-3 and $A_{binary}(r,\theta)$ is a either unity or zero. Fig-4 shows the intensity, i.e. shadow, at the telescope aperture plane for the graded (left) and binary (right) occulter and Fig-4 middle plots a trace through each. In the center, where the telescope resides, the suppression is ~$10^{-10}$ for both graded and binary occulters. The primary difference is that towards the dark hole edges more intensity modulation is evident for the binary occulter whereas the graded occulter intensity is smooth.

### 2.3 Occulter-to-Telescope Propagation

Propagation of the light from the occulter to the telescope aperture is via scalar Fresnel propagation [12]. The field incident on the occulter from a distant source is effectively a plane wave $E_I(x,y) = E_0(\lambda)e^{i\frac{2\pi}{\lambda}(\alpha x + \beta y)}$ where $(\alpha,\beta)$ are small incidence angles with respect to the occulter surface normal, $(x,y)$ are coordinates along the occulter surface and $E_0(\lambda)$ is the field strength at wavelength $\lambda$. An overall phase factor in $z$ has been neglected. The effect of the occulter is to multiply the incident field by a transmission function $T(x,y)$, which can be complex, however herein $T(x,y) = 0$ interior to, and $T(x,y) = 1$ exterior to the occulter. The field propagation from $z = 0$ (occulter plane) to $z$ (telescope aperture plane) is accomplished by the Fresnel integral:

$$E_{II}(x',y';z) = \frac{-i}{\lambda z}\int_{-\infty}^{\infty}\int_{-\infty}^{\infty} T(x,y)E_I(x,y)e^{i\frac{\pi}{\lambda z}[(x-x')^2 + (y-y')^2]}dxdy \qquad (1)$$

where $(x',y')$ are output coordinates at the plane of the telescope aperture.

The transmission function is represented as $T(x,y) = 1 - A(x,y)$, where $A(x,y)$ is the amplitude attenuation function. Using this and changing to polar coordinates by defining $x = r\cos\theta$, $y = r\sin\theta$, $x' - z\alpha = \rho\cos\phi$ and $y' - z\beta = \rho\sin\phi$, better exploits the occulter's symmetry to yield:

$$E_{II}(\rho,\phi;z) = E_0(\lambda)e^{i\frac{2\pi}{\lambda}(\alpha\rho\cos\phi + \beta\rho\sin\phi)}e^{i\frac{\pi}{\lambda}\{\alpha^2+\beta^2\}z}\left\{1 + \frac{i}{\lambda z}e^{i\frac{\pi}{\lambda z}\rho^2}\int_0^{R_{max}}e^{i\frac{\pi}{\lambda z}r^2}rdr\int_0^{2\pi}A(r,\theta)e^{-i\frac{2\pi}{\lambda z}r\rho\cos(\theta-\phi)}d\theta\right\} \qquad (2)$$

where $R_{max}$ is the maximum radius of the occulter's attenuation function.

A planet or star, at differing field angles, is modeled by changing $(\alpha,\beta)$ which shifts the output coordinates by $(\alpha z, \beta z)$, shifting the location of the shadow and tilting the phase with respect to the telescope. The phase of the shadow field is tilted by a plane wave $e^{i\frac{2\pi}{\lambda}(\alpha x' + \beta y')}$ and there is an additional phase shift of the form $\frac{\pi}{\lambda}(\alpha^2 + \beta^2)z$ due to the increase in path length of this tilted beam.

Equation (2) can be evaluated, for a binary occulter, by defining $h(r;\rho,\phi) = \int_0^{2\pi} A(r,\theta)e^{-i\frac{2\pi}{\lambda z}r\rho\cos(\theta-\phi)}d\theta$ and evaluating it as a summation over weighted Bessel functions [7]:

$$h(r;\rho,\phi) = J_0\left(\frac{2\pi}{\lambda z}r\rho\right)N_s\Delta\theta(r) + 4\sum_{k=1}^{\infty}(-1)^k\frac{1}{k}J_{kN_s}\left(\frac{2\pi}{\lambda z}r\rho\right)\cos\left(kN_s\left(\frac{\pi}{2}-\phi\right)\right)\sin\left(\frac{kN_s}{2}\Delta\theta(r)\right) \quad (3)$$

$N_s$ is the number of petals or spokes in the occulter and $\Delta\theta(r)$ is the angle subtended by a single occulter petal vs. radius. Bessel terms within the summation become rapidly smaller with increasing index k and numerically it was found that only ~10 terms were needed to accurately model the occulters shadow field. The recipe for modeling the occulter consists of using quadrature to evaluate the integral over $r$ in equation (2) where $h(r;\rho,\phi)$ is evaluated by summing the series in equation (3) for k=1,…,10. The intensity at the occulter shadow is subsequently given by $|E_{II}|^2$. Numerically a uniform rectilinear output sampling grid of size $N\times N$ spaced by $\Delta x$ is used and equation (2) is numerically evaluated for each output sample point converted from rectilinear to polar coordinates. The output grid is centered on the center of the shadow at $\rho=0$; thus depending on $(\alpha,\beta)$ the telescope aperture shifts within the shadow grid and a tilted phase factor is added. Propagation from the telescope aperture to the focal plane is performed by applying an aperture mask of the form $circ\left(\frac{\sqrt{x'^2+y'^2}}{D}\right) = \begin{cases}1 & \text{for } \sqrt{x'^2+y'^2}\le D/2 \\ 0 & \text{otherwise}\end{cases}$ (D = telescope diameter) to the shadow field i.e. $E'_{II}(\rho,\phi;z) = circ\left(\sqrt{x'^2+y'^2}/D\right)E_{II}(\rho,\phi;z)$ and propagating to the focal plane via FFT techniques:

$$E_{III}(\theta_x,\theta_y) \propto \int_{-\infty}^{+\infty}\int_{-\infty}^{+\infty}circ\left(\frac{\sqrt{x'^2+y'^2}}{D}\right)E_{II}(x',y';z)e^{-i2\pi\left(\frac{\theta_x}{\lambda}x'+\frac{\theta_y}{\lambda}y'\right)}dx'dy' \quad (4)$$

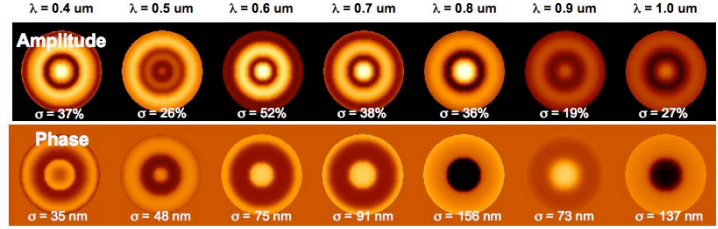

**Figure 5** – *Amplitude & Wavefront Error of Leakage Field within Telescope Aperture*

$(\theta_x,\theta_y)$ are detector coordinates projected on the sky, i.e. angular sky coordinates. The output field from (4) is normalized such that its intensity integral $\iint|E_{III}|^2 d\theta_x d\theta_y$ is equal to the intensity integral over the aperture $\iint_{Aper}|E'_{II}|^2 \rho d\rho d\phi$ to conserve flux.

## 2.4 Baseline TPF-O Performance

The baseline TPF-O architecture is a 50 meter diameter, 16 petal occulter placed at a separation of 72,000 km from a 4 meter aperture telescope that would nominally operate over the spectral band of 0.4 – 1.0 microns. This architecture's performance is shown graphically in Fig-6, and is scalable to other sizes as discussed in section 2.1. Shown on the left of Fig-6 are monochromatic and broadband shadow intensity. The middle top shows plots the center of the shadow's intensity for monochromatic, V-band, and Open-band (0.4 – 1.0 um) and the lower middle plots the intensity averaged across a 4 meter aperture versus wavelength. It is seen that the aperture averaged flux is

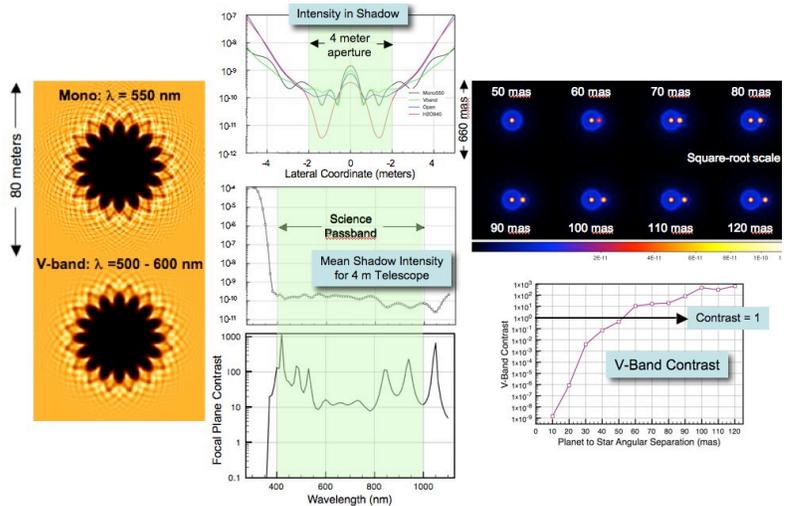

**Figure 6** – *Baseline TPF-O Performance*

~$10^{-10}$ across the entire spectral band of interest. Note that R. Vanderbei optimized the occulter design for this passband and performance is expected to degrade outside this range. The flux averaged across the aperture gives the suppression factor at the aperture but not the focal plane contrast due to the residual suppressed field being brought to focus. This residual field has some small amount of amplitude and wavefront errors across the aperture as shown in Fig-5. The rms amplitude error varies from 19 – 52% rms and the rms wavefront error from 35 to 137 nm rms ($\lambda/14$ to $\lambda/4$) over the spectral band. These errors plus the diffraction from the aperture spread the leakage flux into a leakage point spread function (PSF) as can be seen in the upper right of Fig-6. The amplitude and wavefront errors are primarily radially symmetric and appear like small amounts of defocus and spherical aberration and it is this effect that is responsible for the diffuse rings around the leakage PSF. The contrast of the planet-to-starlight leakage thus varies with location in the focal plane as plotted in the lower right of Fig-6. This contrast exceeds unity for > 52 mas and is >10 at the geometric IWA of 72 mas. Thus the effective IWA is less than the geometric implying it may be possible to detect planets that are inside the geometric footprint of the occulter. The stellar leakage decays as $\sim 1/\theta^3$ where $\theta$ is the angular separation between planet and star, thus the contrast increases with increasing $\theta$.

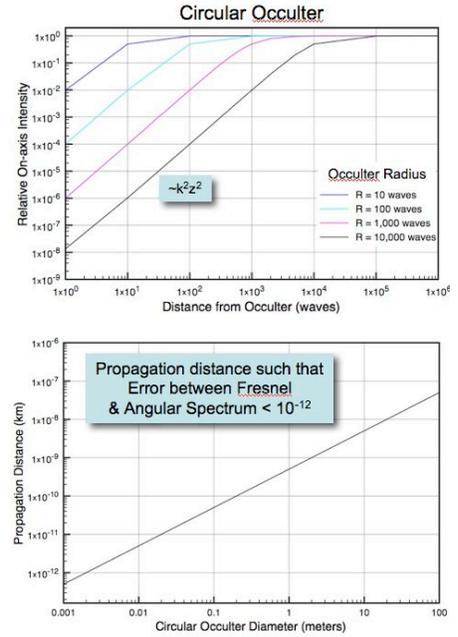

**Figure 7** – *Fresnel Occulter Accuracy*

## 3. SENSITIVITY ANALYSIS

### 3.1 Accuracy of Fresnel Diffraction for Circular Occulter

Is Fresnel propagation accurate to << $10^{-10}$ in intensity within the shadow or is a more rigorous theory required ? This can be assessed for a uniform circular occulter, i.e. without petals, since the shadow field in the Fresnel approximation is given by:

$$E(\rho) = 1 + \frac{i2\pi}{\lambda z} e^{i\frac{\pi}{\lambda z}\rho^2} \int_0^R e^{i\frac{\pi}{\lambda z}r^2} J_0\left(\frac{2\pi r\rho}{\lambda z}\right) r\, dr \qquad (5)$$

and at $E(\rho)\big|_{\rho=0} = 1 + \frac{i2\pi}{\lambda z}\int_0^R e^{i\frac{\pi}{\lambda z}r^2} r\, dr = e^{i\frac{\pi}{\lambda z}r^2}$ yields an intensity of $I(\rho)\big|_{\rho=0} = 1$ independent of distance from the occulter. This appears as an inconsistency since at z = 0 the light would need to diffract into a $90^0$ angle at which point it becomes evanescent and exponentially damped. Thus light would appear where it would be expected to be dark.

Assessing the same problem with more rigorous Rayleigh-Sommerfeld (RS), implemented via angular spectrum [11], which starts with the Helmholtz equation (time harmonic wave equation): $\nabla^2 E(\vec{r}) + k^2 E(\vec{r}) = 0$. Taking its 2D Fourier transform with respect to x, y, but not z, yields a 2nd order ordinary differential equation with derivatives only in the distance variable z: $\{k^2 - 4\pi^2(f_x^2 + f_y^2)\}E(f_x,f_y;z) + \frac{\partial^2}{\partial z^2}E(f_x,f_y;z) = 0$ where $(f_x, f_y)$ are the spatial frequency variables. Eqn-6 has solutions (for planar boundary conditions) of the form $E(f_x,f_y;z) = E(f_x,f_y;0)e^{ikz\sqrt{1-\lambda^2(f_x^2+f_y^2)}}$ where $E(f_x,f_y;0)$ is the 2D Fourier transform of the occulter shape and the complex phasor steps the field's transform a distance z. An inverse 2D Fourier transform of $E(f_x,f_y;z)$ yields the shadow field as $E(f_x,f_y;0) = \delta(f_x,f_y) - RJ_1\left(2\pi R\sqrt{f_x^2+f_y^2}\right)/\sqrt{f_x^2+f_y^2}$ and in polar coordinates $f_x = \rho'\cos\phi$, $f_y = \rho'\sin\phi$ gives the shadow field as a function of $z$ but along $\rho = 0$ as:

$$E(\rho;z)\Big|_{\rho=0} = e^{ikz} - 2\pi R \int_0^\infty J_1(2\pi R\rho')e^{ikz\sqrt{1-\lambda^2\rho'^2}}d\rho' \quad (6)$$

Evaluating (6) in the limit as $z$ tends to zero yields $\lim_{z\to 0} E(\rho;z)\Big|_{\rho=0} = 0$ the expected physical result, i.e. that there is no light directly behind the occulter. For $z \ll \lambda$ and $R \gg \lambda$ the intensity is $I \approx z^2\left\{k^2 + \frac{1}{R^2} - \frac{2k}{R}\sin(kR)\right\} \approx k^2 z^2$. Thus close to the occulter the intensity is proportional to the square of the distance from it. Eqn-6 was numerically evaluated and the intensity plotted versus distance from the occulter for various occulter sizes and the result is plotted in Fig-7 top. The circular occulter intensity quickly asymptotes to unity beyond a few 10,000's of wavelengths of light. Fig-7 bottom plots the propagation distance z such that the absolute error between the Fresnel and RS theory is less than $10^{-12}$ versus occulter diameter. Propagation from an occulter of diameter 50 meters beyond a propagation distance of $3\times 10^{-8}$ km yields a difference between the two theories of $< 10^{-12}$ and beyond this the Fresnel theory is accurate enough.

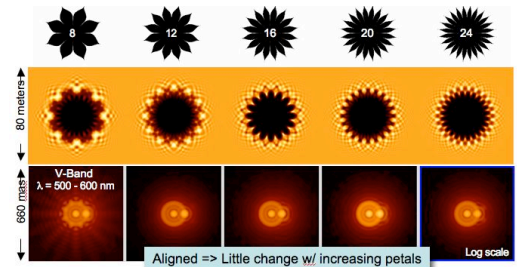

Figure 8 – *Shadow & Planet Images vs. Number of Petals*

### 3.2 Number of Occulter Petals

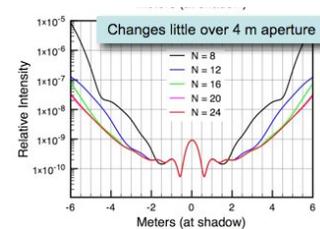

Figure 9 – Shadow and Aperture average Intensity

As was seen in section 2.2 in approximating a continuously graded occulter by a petaled one the number of the petals must be chosen. The effect of increasing number of petals are smoother walls of the shadows intensity as seen in Fig-8 middle and plotted in Fig-9 which show the shadow intensity for petals from 8 – 24 in increments of 4 petals. The shape of the intensity within the center of the shadow, i.e. across the telescope's aperture varies little in the range of +/-2 meters except at the very edge. This edge variation makes the external occulter more sensitive to misalignments. Fig-10 plots the contrast vs. number of petals for a planet at the geometric IWA in the x-direction but with the occulter shifted by 1 meter in X (top) and in Y (bottom) at 4 wavelengths. A contrast of $>10^{10}$ can be met for 14 or more petals over the wavelength range. If the wavelength range is truncated to 0.55 – 1.0 um then 10 petals is acceptable. If contrast is required to be $10^{11}$ over the entire spectral range 16 or more petals are required, and if the spectral range is shortened to 0.55 – 1.0um then 12 or more petals are required. To meet the requirement of an allowable +/1 meter occulter shift, over the spectral range of 0.4 – 1.0 microns, with a contrast of $10^{11}$ at the IWA requires 16 or more petals.

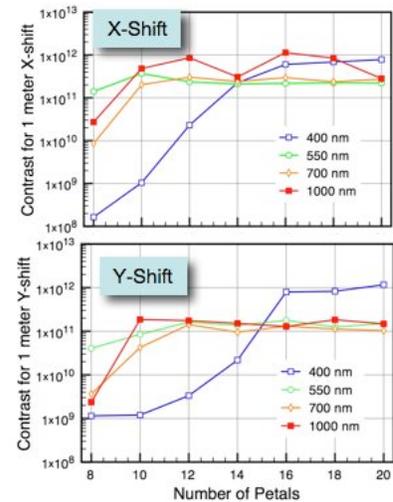

Figure 10 – *Contrast at IWA w/ 1 meter X/Y shift*

### 3.3 Occulter Model with Shape Errors

Manufacturing, deployment, micro-meteorite hits in flight, and deformations of the occulter due to excitations of structural eigenmodes during thrusting and due to reaction/momentum wheels during observations all need to be assessed. Errors of these types are likely to cause the suppression depth of the occulter's shadow field to be less than the design value, i.e. brighter, allowing more star light to leak through to the science focal plane and reducing the contrast. Initial modeling of shape errors and holes errors is done by constructing a perturbed model of the external occulter and propagating from it to the telescope aperture.

Let shape errors in the occulter attenuation function be represented by $\delta A(r,\theta)$ and the full attenuation function be given by $A(r,\theta) + \delta A(r,\theta)$ where $A(r,\theta)$ is the unperturbed attenuation function for which the solution is found via equations (2) and (3). The perturbed shadow is calculated by Fresnel diffraction and coherently added to the unperturbed field, i.e. $E'(\rho,\phi) = E(\rho,\phi) + \delta E(\rho,\phi)$. This approach lends itself well to numerical analysis since the full solution only needs to be solved once.

The perturbing field arises from the shape errors and/or holes, bumps along the edges, or "bites" or pieces missing from the edge and these errors can be thought of as a new source, i.e. a radiator or absorber which coherently interferes with the unperturbed light. In Cartesian coordinates the perturbing field in the shadow is given by:

$$\delta E(x',y') = \frac{i}{\lambda z} \int_{-\infty}^{+\infty}\int_{-\infty}^{+\infty} \delta A(x,y) e^{i\frac{\pi}{\lambda z}\left[(x-x')^2 + (y-y')^2\right]} dx dy \quad (7)$$

$\delta A(x,y)$ is decomposed into a set of M primitive features, with each primitive centered on $(x_k, y_k)$ and of width $s_k$, of the form $\delta A(x,y) = \sum_{k=1}^{M} a_k rect\left\{\frac{x-x_k}{s_k}\right\} rect\left\{\frac{y-y_k}{s_k}\right\}$ where $a_k$ is the attenuation, (for an edge bump, or the gain for a hole or tear) of that decomposed piece of the feature and $rect\left\{\frac{x-x_k}{s_k}\right\} rect\left\{\frac{y-y_k}{s_k}\right\}$ is the shape function of that piece of the feature – e.g. a small tear or hole in the occulter can be decomposed into a set of fine rectangular, triangular or circular functions, i.e. any basis functions of which the 2D Fourier transform is known. The widths of the primitives of the decomposed features are chosen small such that their Fresnel numbers are $s_k^2/\lambda z \ll 1$ implying that the occulter's shadow is in the far field of each of the primitives of the decomposed pieces. Under these assumptions equation (7) can be evaluated to yield for $\delta E(x',y') \approx$

$$\frac{i}{\lambda z} \sum_{k=1}^{M} a_k s_k^2 sinc\left(\frac{s_k}{\lambda z}(x'-x_k)\right) sinc\left(\frac{s_k}{\lambda z}(y'-y_k)\right) e^{i\frac{\pi}{\lambda z}\left[(x'-x_k)^2 + (y'-y_k)^2\right]}$$

(8)

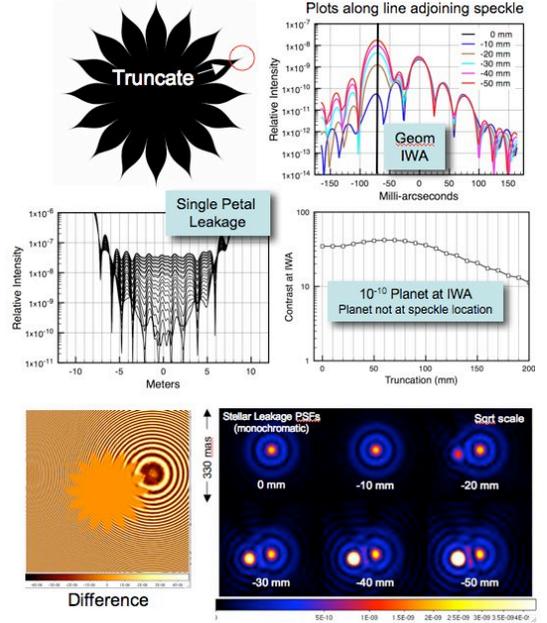

Figure 11 – *Single Petal Truncation Error*

Thus a simple summation over only the region with errors is required, making this approach numerically fast for localized errors. In practice the shape of primitive function matters little under the assumption of operating in the primitive's far field. This is due to these localized occulter features having a slowly varying global effect on the occulter shadow, thus e.g. assuming a single primitive feature width of 1 mm, wavelength of 0.5 microns and distance to occulter of 72,000 km yields a Fresnel number of $2.8 \times 10^{-8}$ which is much less than unity and hence in the far-field. For a 50 meter diameter occulter with a 4 meter aperture the sinc function varies from center-to-edge by 0.99999997969 and thus the sinc product in equation (8) varies from 1 to 0.99999995938 or 4 parts in $10^8$ in field or $1.6 \times 10^{-15}$ in intensity across the aperture due to the 1 mm$^2$ primitive. This is true nearly independent of the shape of the primitive so long as the primitive is small with respect to the far field distance. The amplitude of the field is given by $s_k^2/\lambda z = 2.8 \times 10^{-8}$ and intensity by $7.7 \times 10^{-16}$. Thus a relatively straightforward analysis can be used to treat shape errors, first by calculating the ideal field and coherently combining it with the perturbing field. This analysis was used to treat single petal truncation errors, truncation of all petals and holes within the occulter.

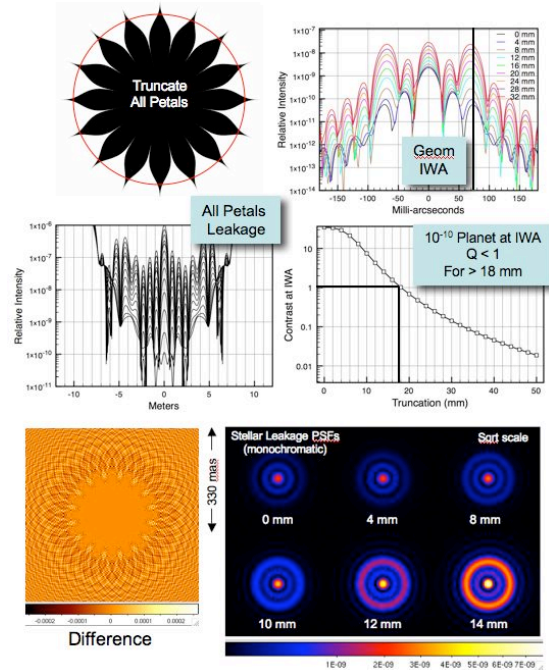

Figure 12 – *All Petals Truncation Error*

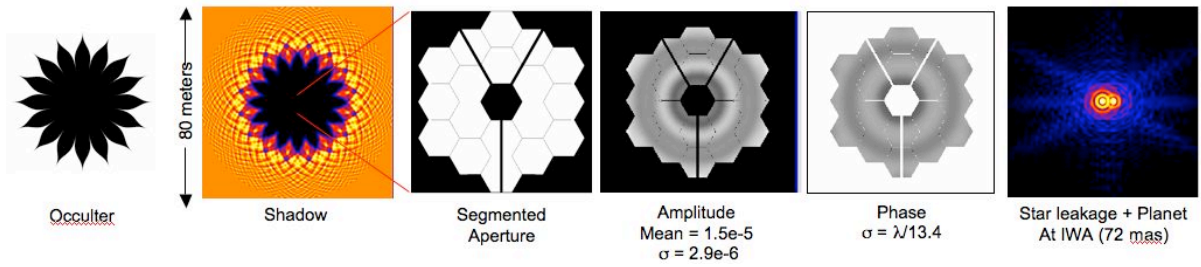

**Figure 13** – *Occulter w/ Segmented Aperture*

### 3.4 Single Petal Truncation Error

Truncation errors for a single petal were model via the analysis in section 3.3. A single petal, red circle upper left in Fig-11, was truncated from 0 to 220 mm in 10 mm increments and propagated to the telescope aperture and through to the focal plane. Fig-11 middle left shows the effect of this truncation within the occulter shadow - the light within the shadow increases with a slight gradient towards the truncated petal. The bottom left right of Fig-11 is the difference between an error free occulter and a truncated petal occulter intensity – a series of concentric rings centered on the truncated petal tip is evident. The bottom right of Fig-11 shows the stellar leakage in the focal plane with increasing truncation; at less than 10mm of truncation no evidence of the truncation visually appears but for >10mm of error a bright localized speckle appears towards the lower left and this speckle increases in brightness with increasing error until it dominates the stellar leakage beyond 30mm. The top right of Fig-11 shows plots through the focal plane in the direction of the speckle. Errors of this type appear like focal plane speckle and can be confused with planets. The middle right of Fig-11 shows the contrast between a planet and the leakage with increasing truncation error for a planet located at the IWA but not at the location of the speckle, i.e. along x=0. This function varies slowly and if the location of the planet and the damaged petal is known a-priori then the occulter could be rotated to move the speckle away from the planet but otherwise an allowable single petal truncation error is ~10 mm or less.

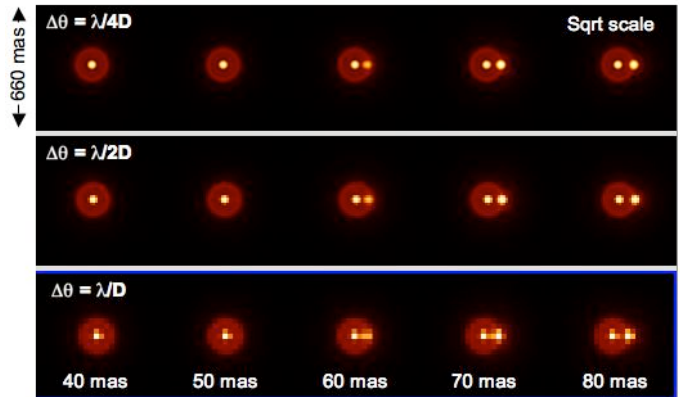

**Figure 14** – *Effect of Sampling*

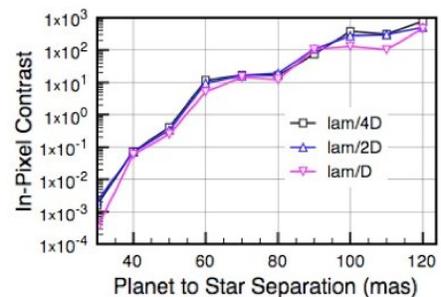

**Figure 15** – *Contrast w/Sampling*

### 3.5 All Petals Truncated Error

All petals were truncated by the same amount to assess the effect and the results shown in Fig-12. The middle left shows the increase in brightness of the occulter shadow intensity with increasing truncation of all the petals. The lower left shows the difference between the error free and the truncated petal occulter's shadow intensity. The right shows the focal plane stellar leakage – a bright ring appears at the IWA and grows with increasing truncation error. Fig-12 upper right shows plots through the focal plane leakage and middle right shows the contrast at the IWA with increasing truncation error. It is seen that the contrast drops from ~30 without any errors to unity at truncation errors of ~18 mm and to hold the contrast at 10 or above requires that all petals be truncated by 8 mm or less.

### 3.6 Filled, Segment and Sparse Aperture Telescopes

The external occulter suppresses the starlight prior to entering the telescope with minimal modification of the planet light. Thus a filled, segmented or even a sparse aperture telescope will effectively collect the light and bring it to focus. The main effect of the telescope aperture shape is the shape of the diffraction in the focal plane. More diffractive structure within the aperture, such as segment boundaries, gives more diffractive spreading decreasing photometric sensitivity. An example is shown in Fig-13 with a telescope consisting of a segmented on-axis primary mirror with a secondary mirror mounted on spider. The image on the far right of Fig-13 shows the focal plane image consisting of leaked starlight and planet. Diffraction flairs are evident in this color stretched image thus slightly reducing the photometric sensitivity. An external occulter system will work with filled, segmented or even sparse aperture systems; thus opening up the trade space for alternative flight architectures.

### 3.7 Focal Plane Sampling

Fig-14 shows visually the effect of different samplings on focal plane images. The top row of Fig-14 shows images of stellar leakage plus planet sampled and spatially integrated over detector pixels that are $\lambda/4D$ projected on the sky for star-planet separations of 40,50,60,70 and 80 mas, while the middle row for $\lambda/2D$ and the bottom row for $\lambda/D$ pixels. The $\lambda/D$ pixels appear visually coarse, however, contrast in a pixel for the 3 samplings are plotted versus angular separation in Fig-15 and do not significantly differ for the 3 cases. The SNR will vary with sampling.

### 3.8 Sensitivities Calculations in Progress

Other sensitivities and errors are still being studied. These include: bumps and bites out of the occulter edges, curved and distorted petal(s), deformation modes of the occulter, sky illumination of the occulter darkside and solar illumination of the occulter edges and straylight within the telescope.

## 4. OCCULTER ACQUISITION AND POINTING

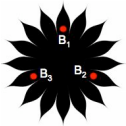

**Figure 16** *Beacons*

Following completion of a given science observation the occulter is moved with thrusters to the new target star and the telescope repointed. After the telescope has acquired and locked onto the target star it needs to acquire and sense the location of the occulter to insure that the contrast is maintained throughout the observation. An initial approach under study is a two-stage process consisting of using up to 3 small laser beacons (diode lasers) mounted on the occulter as shown in Fig-16. Each beacon would be a diode laser of ~1mm² aperture and as an example we use $\lambda$ = 635nm and emitting ~1 milliwatt of power each. Each of the lasers could be pulse code modulated at a different temporal frequency. The focal plane of the science instrument would be separated with beamsplitters into 3 channels: a science channel, a pupil imaging channel and a coarse acquisition channel as shown schematically in Fig-17.

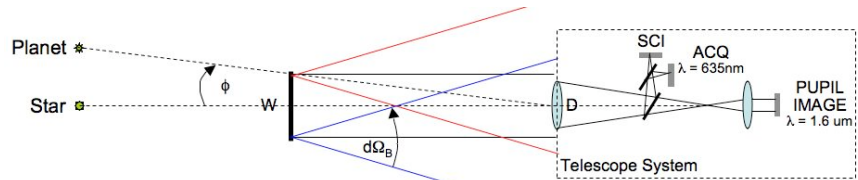

**Figure 17**- *Occulter Acquisition*

The coarse acquisition channel would acquire one or more spots from the diode lasers. Each laser's beam would diverge from the occulter into an angle of $\theta_F \approx 2\lambda/\pi w_0 = 0.4$ milli-radians (83 arcseconds). Thus if the occulter was laterally shifted by ½ this angle, i.e. a distance of +/- 0.2 milli-rads * 72,000km = +/- 29 km it's laser beacons would be seen and focused by the telescope. The field-of-view of the telescope plus coarse acquisition sensor should be at least 83 arcseconds. If the 3 laser beacons are placed on the dark side of the occulter at 60% of the occulter radius, i.e. at 15 meters from its center then 26 meters would separate them all from each other. At 72,000 km this is an angular separation of the spots by 2.3 $\lambda/D$ for a 4 meter telescope implying the spots are easily discriminated in the coarse acquisition focal plane. Providing 3 laser beacons in addition to supplying redundancy also allows the lasers to be slightly shifted in angle, i.e. off normal to the occulter giving effectively a larger capture range.

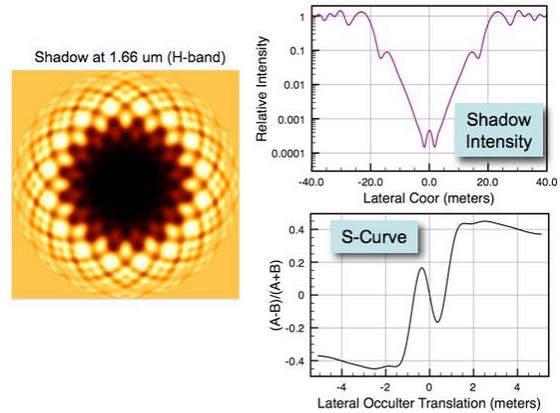

**Figure 18** – *Occulter Shadow at λ = 1.6 um*

The attitude control system on the occulter would insure that the normal to the occulter surface was aligned to the target star to +/-5 arcseconds. The collected laser power is ~ $P_L \cdot \Omega_{Tel}/\Omega_{Beam} \approx 10^{-8} P_L$ where $P_L$ is the power of the laser, $\Omega_{Tel}$ is the solid angle subtended by the telescope as seen from the occulter and $\Omega_{Beam}$ is the solid angle that the laser beam diverges into. A 1 milliwatt laser gives approximately $10^6$ photons/second collected by a 4m telescope at 72,000 km yielding a SNR of ~1000 in 1 second. Use of 4096 x 4096 focal-plane array mapped to a capture range of +/-83 arcseconds gives $1.3\lambda/D$ sampling and a simple centroiding algorithm could resolve the occulter location to <¼ of this or at the occulter to a resolution of +/-3.5 meters.

Following the coarse acquisition the pupil imaging channel would be used to sense the occulter's lateral location within 1 meter of its ideal location along the line of sight to the target star. The pupil imaging channel images a region of the occulter's shadow truncated by the telescope aperture, i.e. an image of the telescopes entrance pupil, onto a near-IR camera at 1.6 microns. Lateral shifts of the occulter will change the pattern and flux on the pupil imaging camera and using a wavelength outside the spectral band over which the occulter design has been optimized results in an increase in flux as shown in upper right of Fig-18 – the depth of the shadow is only ~0.0002 in H-band as opposed to ~$10^{-10}$ in V-band within the spectral band optimized shadow. Once the occulter's shadow walks entirely off the telescope aperture only a uniform illumination of the pupil will result. Fig-18 lower right shows the x-direction error signal, aka S-curve, as a function of lateral translation of the occulter. The error signal is the integral of the left half of the pupil image differenced from the integral of the right half divided by the sum of the flux in the entire pupil. The double bump value in the S-curve center comes from the "bump" in the center of the shadow intensity (Fig-18 top right) and varies with wavelength. Other choices for the fine acquisition channel wavelength will make the S-curve single valued. It is this S-curve, in both x- and y-directions, along with the starlight suppressed science image which can be used to fine sense the location of the occulter and feedback to the occulter's control system to maintain its position throughout an observation.

**Table 2:** *TPF-O Baseline System and Sensitivities*

| Occulter + Telescope System | | |
|---|---|---|
| Occulter Diameter: W = | 50 | meters |
| Telescope Diameter: D = | 4 | meters |
| Separation: z = | 72,000 | km |
| Wavelength Range: Δλ = | 0.4 - 1.0 | um |
| Fresnel # @ λ = 0.5: $F_N$ = | 69.4 | |
| Number of Petals: N = | 16 | |
| Occulter form: | Vanderbei | |
| Geometric IWA: | 72 | mas |
| Effective IWA: | 52 | mas |
| Focal plane contrast: > | 10 | |
| **Sensitivities** | | |
| Occulter Holes: Area < | 3 | cm$^2$ |
| Single Petal Error: < | 10 | mm |
| All Petal Truncation Error: < | 8 | mm |
| Misalignment Sensing: < | 10 | cm |
| Misalignment Control: < | +/- 1 | meter |
| **Coarse Acquisition (Laser Beacons)** | | |
| Number of laser beacons = | 3 | |
| Wavelength: λ = | 0.635 | um |
| Beacon aperture: $w_0$ = | 1 | mm |
| Focal plane: | 4069 x 4096 | pixels |
| Sampling: Δx = | 33 | mas |
| Capture range: | +/- 29 | km |
| Capture resolution: | +/- 3.5 | meters |
| Time to SNR 1000: | 1 | second |
| **Fine acquisition (Pupil Imaging)** | | |
| Wavelength: λ = | 1.66 | um (H-band) |
| Fine acquisition focal plane: | 1024 x 1024 | pixels |
| Sampling at occulter: Δx = | 3.9 | mm |
| Capture range: | +/- 8 | meters |
| Capture resolution: | 8 | mm |

## 5. SUMMARY AND CONCLUSIONS

Table 2 lists the baseline TPF-O system and a partial list of sensitivities. An external occulter coronagraph has the advantage that the starlight is suppressed before entering the telescope thereby levying no stringent optical requirements on the telescope. This approach has no outer working angle that usually arises from a deformable mirror which corrects speckle out to specific outer working angle on the sky. The inner working angle is nearly independent of wavelength over the optimized spectral band. It works with filled, segmented or even sparse or interferometric aperture approaches, and can be combined, in a hybrid or cascaded fashion, with other approaches either to give a useful margin in the design or to potentially relax requirements on both the external occulter and internal coronagraphic scheme. Acquisition sensing, both coarse and fine, of the occulter is relatively straightforward and imposes no stressing requirements. No active wavefront, amplitude nor polarization sensing or control are required at stringent levels. Some wavefront control, akin to JWST, may be required within the telescope. The system is inherently high throughput since the planet is only suppressed by the occulter if it falls within the outer edge of the occulter. The as-designed occulter has broad spectral response from 0.4 – 1.0 microns. It does require two spacecraft with formation flying control to < 1meter lateral to the line of sight and ~200km in the z-direction. This approach for direct terrestrial planet detection and characterization holds sufficient promise to warrant further investigation.